\documentclass[a4paper,11pt]{article}
\usepackage{pos}
\usepackage{slashed}
\usepackage{subfigure}

\title{Hybrid Renormalization for Baryon Distribution\\ Amplitudes from Lattice QCD in LaMET}

\author*[a]{Mu-Hua Zhang}

\affiliation[a]{Tsung-Dao Lee Institute \& State Key Laboratory of Dark Matter Physics, Key Laboratory for Particle Astrophysics and Cosmology (MOE), Shanghai Key Laboratory for Particle Physics and Cosmology, School of Physics and Astronomy, Shanghai Jiao Tong University,\\
Shanghai 200240, China}

\emailAdd{muhuazhang@sjtu.edu.cn}

\abstract{In our recent work~\cite{LatticePartonCollaborationLPC:2025vhd} on lattice QCD calculation of the baryon leading-twist LCDAs within the framework of LaMET, a novel hybrid renormalization scheme is implemented for octet baryon quasi-DAs, yielding reliable results across both perturbative and non-perturbative regions.

The numerical simulations are performed using CLQCD ensembles with $N_f = 2+1$ stout-smeared clover fermions and a Symanzik-improved gauge action. Calculations are carried out at three lattice spacings, $a = {0.052, 0.077, 0.105}$ fm. After renormalization, the linear divergences inherent in quasi-DAs are effectively removed, leading to smooth and well-behaved continuum coordinate-space distributions. These results demonstrate the viability of hybrid renormalization frameworks for light-baryon quasi-DAs and provide a robust foundation for future LaMET-based determinations of baryon LCDAs.}

\FullConference{
The 26th international symposium on spin physics (SPIN2025)\\
21–26 September, 2025\\
Qingdao (Tsingtao), Shandong Province, China\\}

\begin{document}
\maketitle

\section{Introduction}
Light-cone distribution amplitudes (LCDAs), which describe the longitudinal momentum distribution of partons inside hadrons at high energies, are fundamental non-perturbative inputs for understanding hadron structure and exclusive processes. With the first direct observation of CP violation in a baryonic system~\cite{LHCb:2025ray}, the demand for precise theoretical predictions has been significantly intensified. In this context, an accurate determination of baryon LCDAs is urgently needed.

Since the asymptotic forms were first proposed in 1983~\cite{Chernyak:1983ej} and the Chernyak–Ogloblin–Zhitnitsky (COZ) model~\cite{Chernyak:1987nu} was formulated within the QCD sum-rule framework, the study of baryon LCDAs has remained relatively underdeveloped. Because LCDAs are defined through light-cone correlators, they are not directly accessible in Euclidean lattice. While lattice calculations of the lowest moments of octet-baryon LCDAs have been performed using the operator product expansion (OPE)~\cite{Bali:2015ykx,RQCD:2019hps,Bali:2024oxg}, these results remain insufficient for detailed phenomenological applications.

The Large-Momentum Effective Theory (LaMET)~\cite{Ji:2013dva,Ji:2014gla} provides a new framework to compute the $x$-dependence of parton distributions and distribution amplitudes from lattice QCD, and has been successfully applied in a variety of contexts, including the light and heavy meson LCDAs~\cite{LatticeParton:2020uhz,Hua:2020gnw,LatticeParton:2022zqc,LatticePartonLPC:2022eev,LatticeParton:2024zko,Tan:2025ofx}.

In this report, I will present our recent progress on the lattice determination of light baryon LCDAs~\cite{LatticeParton:2024vck,LatticePartonCollaborationLPC:2025vhd}, with particular emphasis on a novel hybrid renormalization scheme designed to eliminate the linear divergences appearing in lattice matrix elements, which lays a solid foundation for future precision calculations.

\section{Quasi Distribution Amplitudes for Light Baryons} \label{sec:framework}

\subsection{Definitions of LCDAs and quasi-DAs for light baryons}

\begin{figure*}[htbp]
\centering
\subfigure[LCDAs in coordinate space.]{
    \centering
    \includegraphics[scale=0.11]{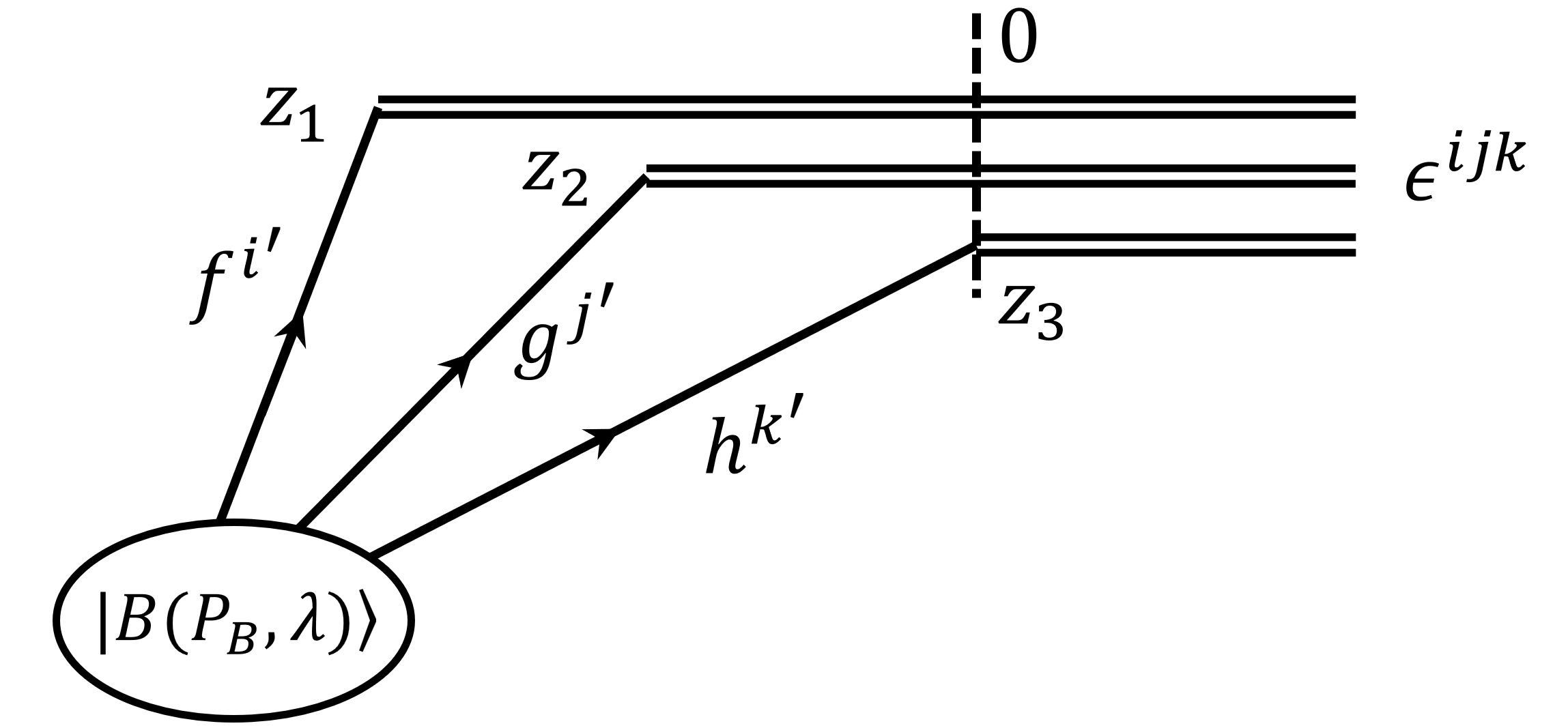}
    }
\vspace{0.0cm} 
\subfigure[LCDAs in momentum space.]{
    \centering
    \includegraphics[scale=0.09]{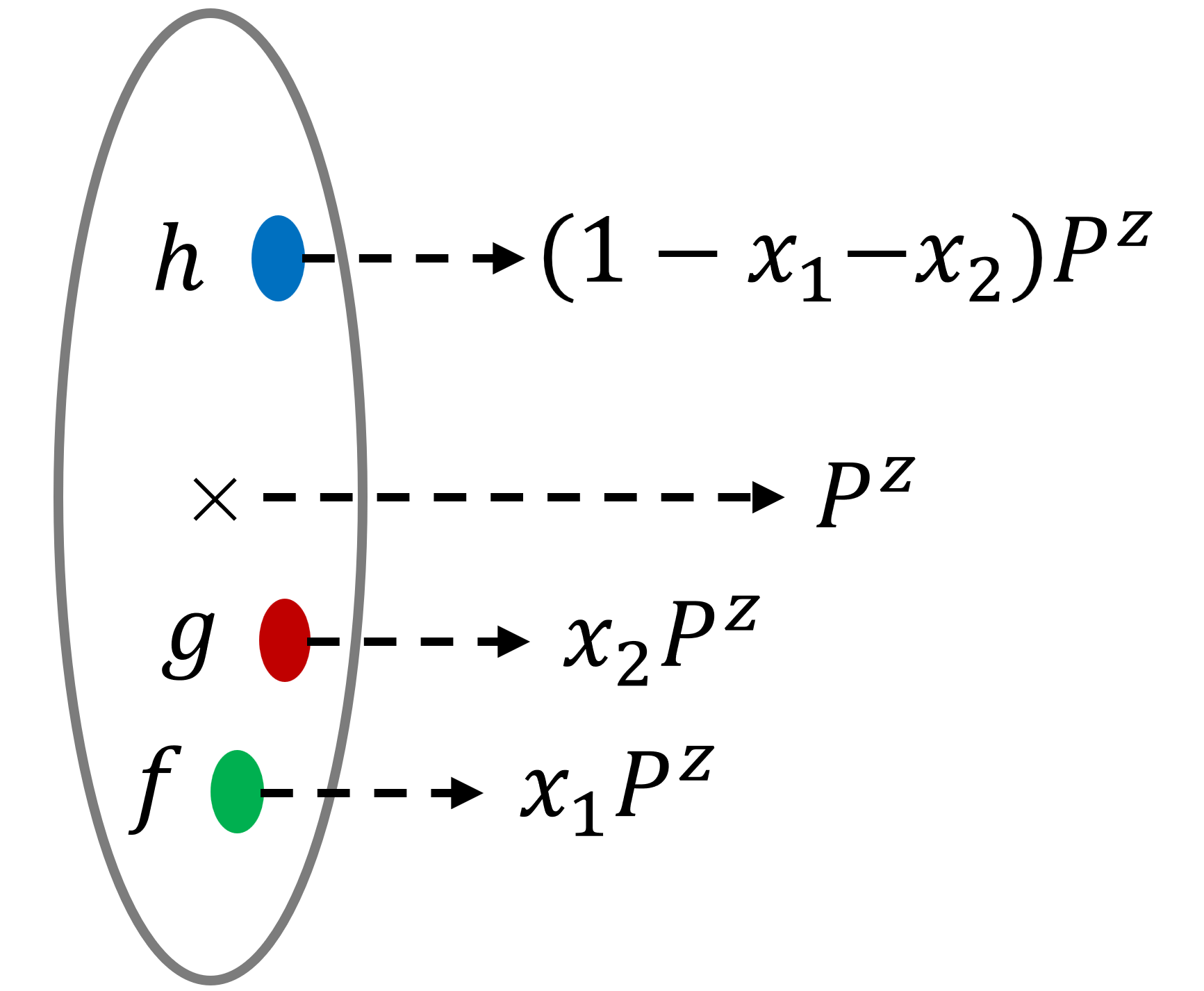}
    }
\caption{The structure of the baryon LCDAs~\cite{LatticePartonCollaborationLPC:2025vhd}.}
\label{fig:structure}
\end{figure*}

The baryon light-cone distribution amplitudes (LCDAs) in coordinate space are defined as the baryon-to-vacuum matrix elements of light-cone-separated non-local operators~\cite{Braun:1999te,Han:2024ucv,Deng:2023csv,Zeng:2025wpd,Shi:2026mjb}: 
\begin{equation}
\begin{split}
\epsilon^{ijk} \langle 0 | 
f_{\alpha}^{i'}(z_1 n) W^{i'i}(z_1 n, z_0 n) g_{\beta}^{j'}(z_2 n) W^{j'j}(z_2 n, z_0 n) h_{\gamma}^{k'}(z_3 n) W^{k'k}(z_3 n, z_0 n) | B(P_B) \rangle,
\label{eq:baryon_matrix}
\end{split}
\end{equation}
where $| B(P_B) \rangle$ represents a baryon state with momentum $P_B^\mu=P_B^{+} \bar n^\mu = (P_B^z,0,0,P_B^z)$, $\alpha,\beta,\gamma$ are Dirac indices, $i^{(\prime)},j^{(\prime)},k^{(\prime)}$ are color indices in fundamental representation, and $f,g,h$ denote the valance quark fields in the baryon state. 
Light-like Wilson lines $ W^{ij} $ connect quark fields on different locations to preserve gauge invariance. Two light-cone unit vectors are $n^\mu=(1,0,0,-1)/\sqrt{2}$ and $\bar n^\mu=(1,0,0,1)/\sqrt{2}$. For simplification we usually set $z_3=0$, as shown in left panel of Fig.~\ref{fig:structure}.

The LCDAs for octet baryons in Eq.~\eqref{eq:baryon_matrix} can be decomposed into 3 functions at leading twist:
\begin{equation}
\begin{split}
&\langle 0 | f_{\alpha}(z_1 n) g_{\beta}(z_2 n) h_{\gamma}(z_3 n) | B(P_B) \rangle \\
=& \frac{1}{4} f^B \Big[ (\slashed{P}_B C)_{\alpha \beta} (\gamma_5 u_B)_{\gamma} \Phi_V^B (z_i n \cdot P_B) + (\slashed{P}_B \gamma_5 C)_{\alpha \beta} (u_B)_{\gamma} \Phi_A^B (z_i n \cdot P_B) \Big] \\
&+ \frac{1}{4} f_T^B (i \sigma_{\mu \nu} P_B^{\nu} C)_{\alpha \beta} (\gamma^{\mu} \gamma_5 u_B)_{\gamma} \Phi_T^B (z_i n \cdot P_B),
\end{split}
\end{equation}
where \( C \equiv i \gamma^2 \gamma^0 \) is the charge conjugation matrix and \( u_B \) stands for the baryon spinor.

The LCDAs in momentum space can be obtained through Fourier transformation:
\begin{equation}
\begin{split}
\phi_{V/A/T}(x_1,x_2) &= \int \frac{P_B^+ dz_1 }{2\pi} \int \frac{P_B^+ dz_2 }{2\pi} \ {\rm e}^{i(x_1z_1+x_2z_2)P_B^+} \Phi_{V/A/T}^B(z_1 n,z_2 n),
\end{split}
\end{equation}
where $x_1,x_2$ are longitudinal momentum fractions carried by $f,g$ quarks, as shown in right panel of Fig.~\ref{fig:structure}. In this work, we focus on the A-term of the $\Lambda$ baryon, as it has non-vanishing local limit.
    
The  proposal of LaMET is that instead of directly calculating the LCDAs, we could calculate the corresponding equal-time correlators at large momentum, known as quasi-DAs, then matching to LCDAs through a factorization. In Euclidean space, the A-term quasi-DA is defined as:
\begin{equation}
\begin{split}
{\widetilde{\Phi}_A}^{B}(z_1, z_2, z_3, P_B^{z}) f^B P_B^{z} u_B &= \left\langle 0 \left| f^{\rm T}(z_1 n_z)(C \gamma_5 \gamma^\nu) g(z_2 n_z) h(z_3 n_z) \right| B(P_B) \right\rangle,
\label{eq:quasiDA_terms}
\end{split}
\end{equation}
where $ n_z $ is a equal-time unit vector along the direction of hadron momentum $ P_B^z $.

\section{Simulation Setup}\label{sec:setup}

\subsection{Lattice setup}
We employ ensembles with $N_f=2+1$ stout smeared clover
fermions and a Symanzik gauge action, generated by the CLQCD collaboration~\cite{CLQCD:2023sdb}. To achieve the proper renormalization scheme while controlling the discretization effects, we use ensembles with 3 different lattice spacings $a=\{0.105,0.077,0.052\}$ fm. More details are summarized in Table~\ref{tab:ensembles}.

\begin{table}[h]
\centering
\caption{Ensembles and lattice setup used in simulation.}
\label{tab:ensembles}
\begin{tabular}{c c c c c c}
        \hline\hline
        Ensembles & $a$ (fm) & $m_\pi$ (MeV) & Volume & $N_{\rm cfg} \times N_{\rm src}$ & $P^z$ (GeV)  \\
         \hline
        C24P29 & 0.1052 & 292.3 & $24\times72 $ & $864 \times 4$ & 0, 0.49, 1.96 \\
        F32P30 & 0.0775 & 300.4 & $32\times96 $ & $777 \times 4$ & 0, 0.50, 2.00 \\
        H48P32 & 0.0520 & 316.6 & $48\times144$ & $550 \times 6$ & 0, 0.50, 1.98 \\
        \hline\hline
    \end{tabular}
\end{table}

\subsection{Operator definitions and extraction of matrix elements}
\label{subsec:Interpolators and projection operators}
In lattice QCD, the matrix elements of quasi-DAs in Eq.~\eqref{eq:quasiDA_terms} can be extracted through a reduction procedure of the corresponding 2-point correlation functions, defined as: 
\begin{equation}
\begin{split}
C_2(t,\vec{P}, z_1,z_2)&=\int d^3x {\rm e}^{-i\vec{P}\cdot \vec{x}}\langle O_{\rm snk}(\vec{x},t;z_1,z_2)_{\gamma} {\bar O}_{\rm src}(0,0;0,0)_{\gamma'}T^{\gamma'\gamma} \rangle.
\label{eq:2pt_definition}
\end{split}
\end{equation}
Here the sink operator $O_{Sink}(\vec{x},t;z_1,z_2)_\gamma$ is determined by the structure that we aim to calculate:
\begin{equation}
\begin{split}
O^{A}_{\rm snk}(\vec{x},t;z_1,z_2)_\gamma = \epsilon_{ijk} f_{\alpha}^{i}(\vec{x}+z_1n_z,t) (C \gamma^t)_{\alpha\beta} g_\beta^{j}(\vec{x}+z_2n_z,t) h^{k'}_{\gamma}(\vec{x},t),
\end{split}
\end{equation}
while for source operator, following the strategy of Ref.~\cite{Zhang:2025hyo}, we adopt a kinematically enhanced form that is constructed to better overlap with the leading Fock state in the boosted frame:
\begin{align} \label{eq:source_boost_p_mod}
\begin{split}
O^\Lambda_{\rm src} &= \frac{1}{\sqrt{6}}\left[ 2(u^{\rm T}C\gamma_5\gamma^t d)s + (u^{\rm T}C\gamma_5\gamma^t s)d + (s^{\rm T}C\gamma_5\gamma^t d)u \right].\\
\end{split}
\end{align}
And for projector $T^{\gamma'\gamma}$, we choose the combination $\gamma^t+\gamma^z$, expecting to project out the leading twist contribution at large momenta.

To improve data quality at large momenta and large $z$-separations, we also employ momentum smeared point source~\cite{Bali:2016lva}, and apply single-step HYP smearing~\cite{Hasenfratz:2001hp,DeGrand:2002vu} to the Wilson lines.

Following the parameterized form of the 2-point function:
\begin{align}
\begin{split}
& C^{\rm norm}_2(t,P^z;z_1,z_2) = \frac{C_2(t,P^z;z_1,z_2)}{C_2(t,P^z;0,0)} = \tilde\Phi(z_1,z_2,P^z)\left(1 + A e^{-\Delta E t}\right),
\label{eq:two_state_fit}
\end{split}
\end{align}
the parameters $A$ and $ \Delta E$ describe the excited states. To suppress the excited-state contamination and reliably extract the matrix elements, we perform a two-state fit to the 2-point functions, not with fixed fit range but implementing the model average procedure introduced in Ref.~\cite{Jay:2020jkz}.

\section{Hybrid Renormalization}\label{sec:frame_hy}

\begin{figure*}[htbp]
\centering
\subfigure[$\Lambda$ bare quasi-DA, $z_1 =0$]{
    \centering
    \includegraphics[scale=0.32]{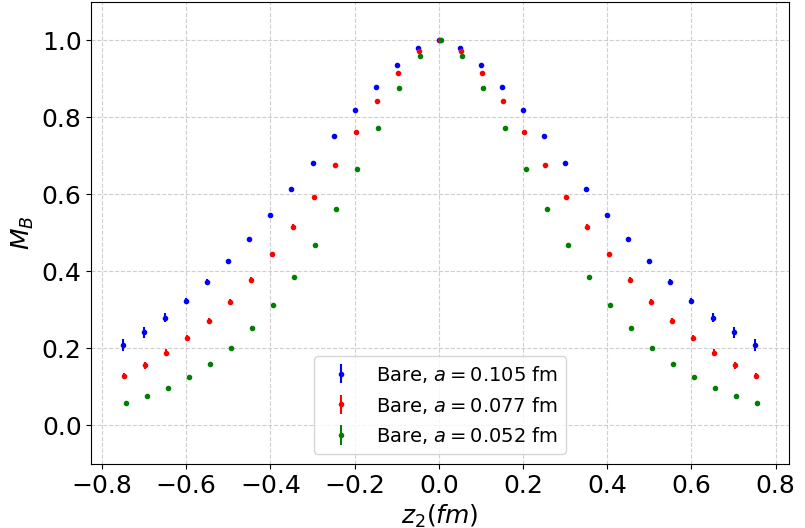}
    }
\vspace{0.0cm} 
\subfigure[$1/a$ dependence in logarithmic scale]{
    \centering
    \includegraphics[scale=0.27]{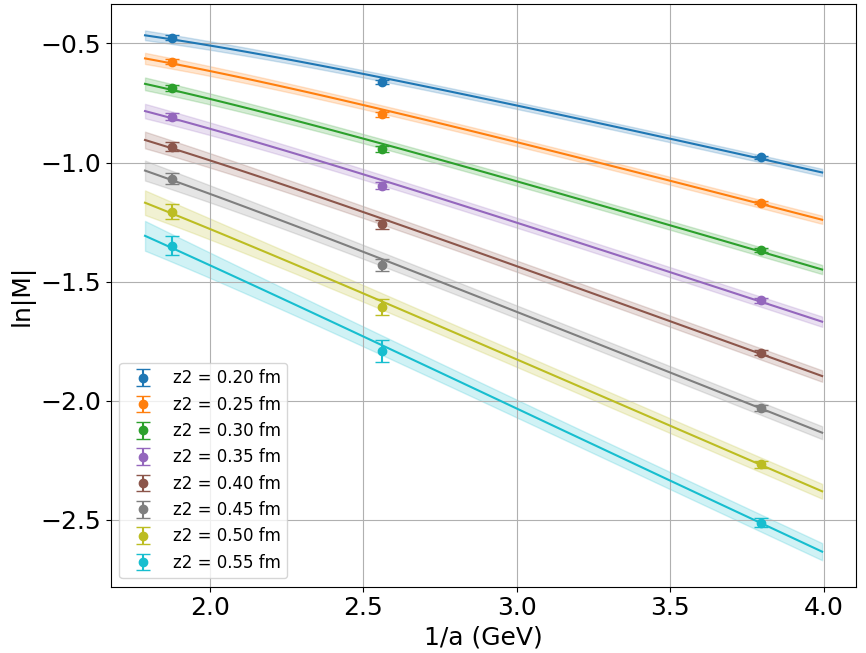}
    }
\caption{Bare 0-momentum quasi-DAs for $\Lambda$ from 3 different lattice spacings~\cite{LatticePartonCollaborationLPC:2025vhd}.}
\label{fig:Normed_Bare}
\end{figure*}

The matrix elements calculated from lattice suffer from discretization effects and divergences that depend on lattice spacings. In left panel of Fig.~\ref{fig:Normed_Bare}, we show the bare quasi-DA of the $\Lambda$ baryon calculated from 3 different spacings. There are significant differences among the results obtained at different spacings, far exceeding the expectations of discretization effects. In right panel of Fig.~\ref{fig:Normed_Bare}, the linear behavior is shown more clear in logarithmic scale. Thus we need a proper renormalization scheme to eliminate the linear divergences and to ensure reliable matching to the $\rm \overline{MS}$ scheme.

\subsection{Review of ratio and self-renormalization schemes}
The Ratio scheme is simply divide the matrix element by its 0-momentum counterpart:
\begin{equation}
\begin{split}
    \hat M_{\rm Ratio}(z_1,z_2;P^z,a) = \frac{\hat M(z_1,z_2;P^z,a)}{\hat M(z_1,z_2;0,a)}.
\end{split}
\end{equation}
Since the ratio scheme is an OPE-based strategy, it could only work well at short range of $z_i$.

When considering the long-range behavior of quasi-DAs, according to the self-renormalization scheme~\cite{LatticePartonLPC:2021gpi}, the divergencies and discrete effects in bare matrix elements can be parameterized as:
\begin{equation}
\begin{split}
\hat M(z_1, z_2, P^z, a) =& \exp \bigg[ \frac{k}{a \ln (a \Lambda_{\rm QCD})}  \tilde{z} + g(z_1,z_2,P^z) +f(z_1, z_2) a^2 \\
& +\frac{\gamma_0}{b_0} \ln \frac{\ln (1 /a \Lambda_{\rm QCD})}{\ln (\mu / \Lambda_{\rm \overline{MS}})} + \ln \bigg(1+\frac{d}{\ln (a \Lambda_{\rm QCD})}\bigg) \bigg], \label{Eq:SelfMz}
\end{split}
\end{equation}
where parameter $k$ describes the linear divergence, $f(z_1,z_2)a^2$ is the discretization effect. $\tilde{z}$ is the effective length of Wilson lines~\cite{Han:2023xbl} defined as $\tilde{z} = \left\{\begin{array}{ll}|z_1-z_2|, & \quad z_1 z_2 < 0 \\ {\rm max}\left(|z_1|,|z_2|\right), & \quad z_1 z_2 \geq 0 \end{array} \right.$.

The quantity $g(z_1,z_2,P^z)$ contains the effects from non-perturbative physics and the mass renormalization parameter~\cite{Ji:1995tm,Zhang:2023bxs}:
\begin{equation}
\begin{split}
g(z_1,z_2,P^z) = \ln \hat M_{\rm \overline{MS}}\left(z_1, z_2,P^z, \mu\right) + m_0 \tilde{z} ,\label{Eq:Selfgz}
\end{split}
\end{equation}
here $\hat M_{\rm \overline{MS}}(z_1, z_2,P^z, \mu)$ represents the matrix element in $\rm \overline{MS}$ scheme, which should be consistent with the one-loop perturbative result $\hat M_p(z_1, z_2,0, \mu)$ given in Ref.~\cite{Han:2023xbl} at short distances.

All the parameters in Eq.~\eqref{Eq:SelfMz} can be extracted from lattice data with perturbative result, through a 2-steps fitting we have demonstrated in Ref.~\cite{LatticePartonCollaborationLPC:2025vhd}, and the renormalized lattice matrix elements $\hat M_{\rm \overline{MS}}(z_1, z_2,P^z, \mu)$ in $\overline{\rm MS}$ scheme can be obtained eventually.

\subsection{Implementation of hybrid renormalization for baryon quasi-DAs}
\label{subsec:Hybrid renormalization}

In our work, we implemented the hybrid renormalization method~\cite{Ji:2020brr}. It is a well-defined scheme for subtracting UV divergences without introducing extra IR effects. The basic idea is to combine the ratio and self-renormalization, adopting different schemes in different regions.

\begin{figure}[htbp]
    \centering
    \includegraphics[width=0.32\textwidth]{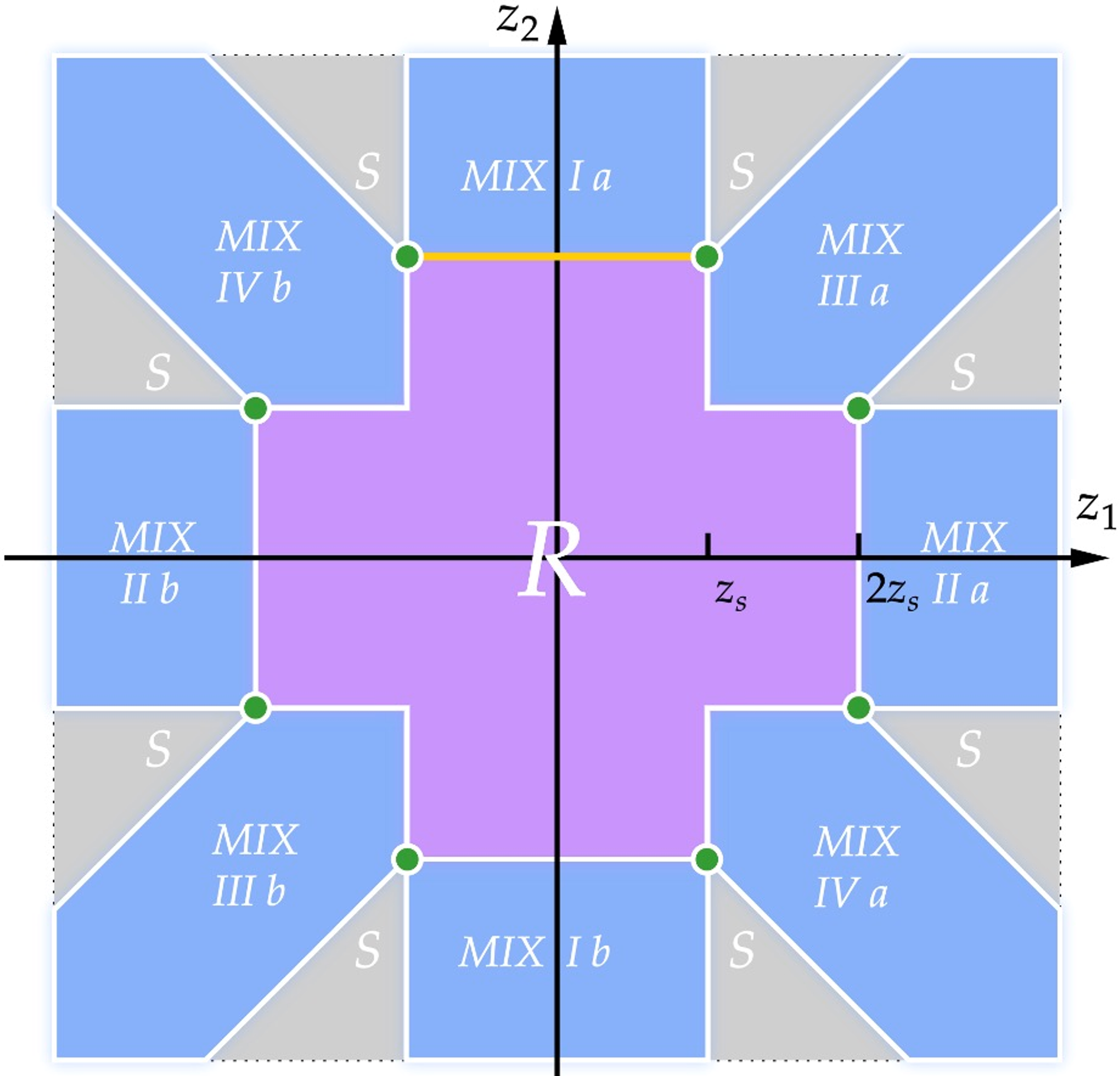}
    \caption{Range division for hybrid renormalization on $z_1$-$z_2$ plane~\cite{LatticePartonCollaborationLPC:2025vhd}.}
    \label{pic:Renorm}
\end{figure}

For the case of baryon DAs, it is rather complicated to implement because we work on a 2-dimensional plane with 3 distinct distances $z_1$, $z_2$ and $z_1-z_2$~\cite{Han:2023xbl,Han:2024ucv,Zhang:2025npd}. We first partition and categorize the $z_1$-$z_2$ plane as illustrated in Fig.~\ref{pic:Renorm}. Here, we choose $a \ll z_s = 0.20 {\rm\ fm} \ll 1/\Lambda_{\rm QCD}$.

\begin{figure*}[htbp]
\centering
\subfigure[\ Bare result]{
    \centering
    \includegraphics[scale=0.27]{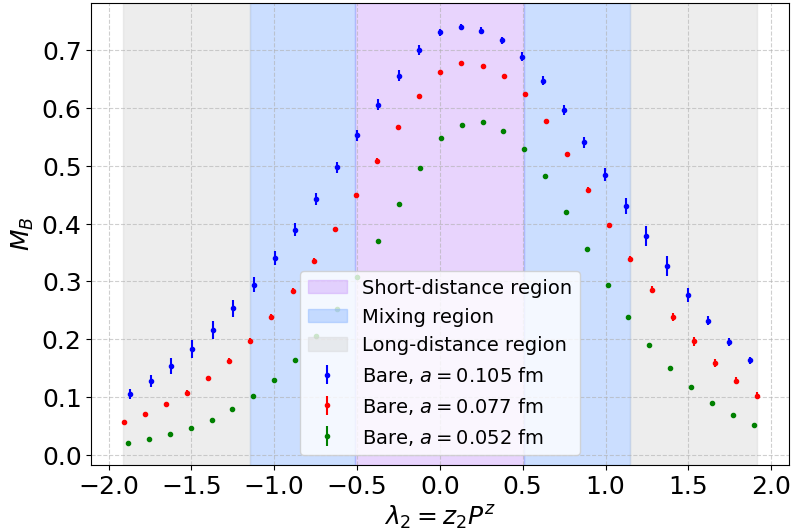}
    }
\vspace{0.0cm} 
\subfigure[\ Hybrid scheme result]{
    \centering
    \includegraphics[scale=0.27]{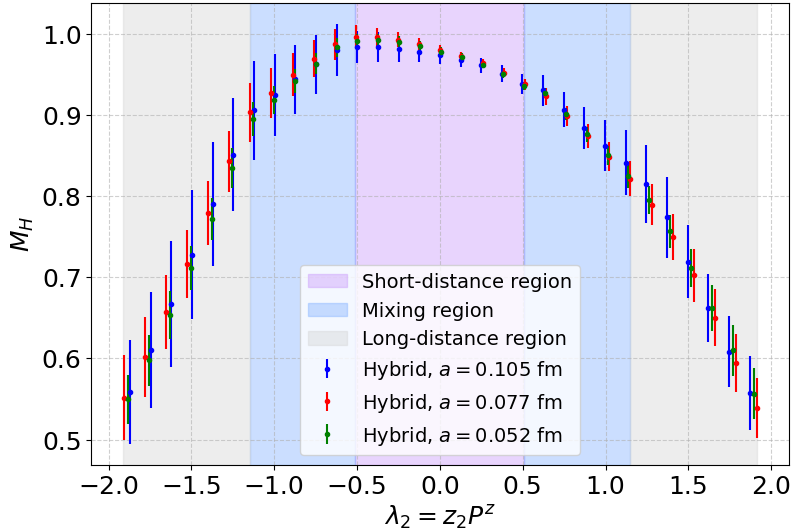}
    }
\vspace{0.0cm} 
\subfigure[\ Ratio scheme result]{
    \centering
    \includegraphics[scale=0.27]{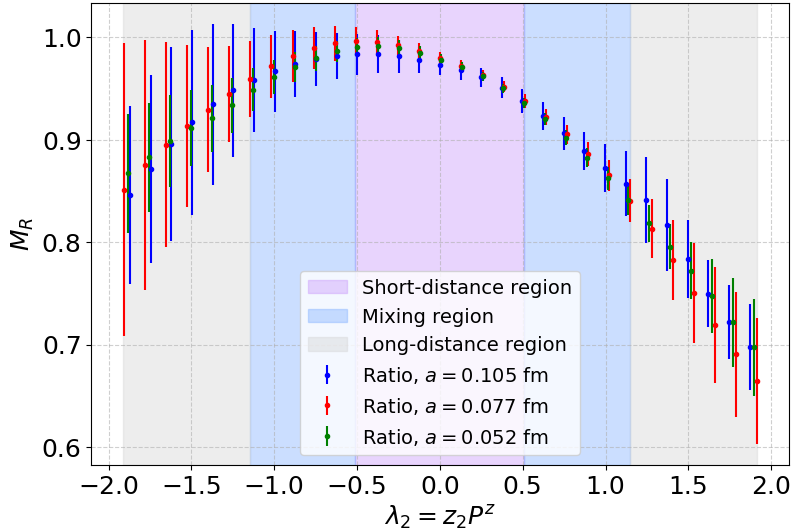}
    }
\vspace{0.0cm} 
\subfigure[\ Self-renormalization result]{
    \centering
    \includegraphics[scale=0.27]{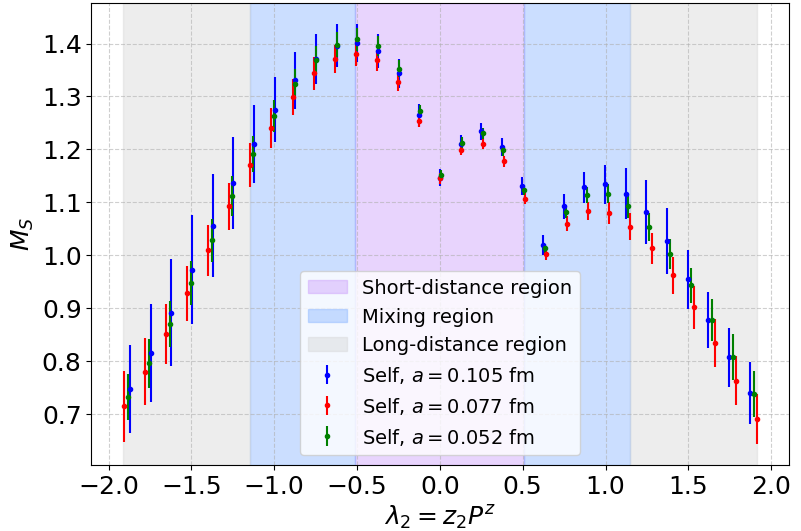}
    }
\caption{Bare, hybrid, ratio \& self renormalization scheme results of $\Lambda$ quasi-DAs at $P^z=0.5$ GeV~\cite{LatticePartonCollaborationLPC:2025vhd}.}
\label{fig:schemes_lambda_pz1}
\end{figure*}

The details and difficulties have been discussed in Ref.~\cite{Han:2023xbl,Han:2024ucv,LatticePartonCollaborationLPC:2025vhd,Zhang:2025npd}, and here I may directly show the comparison of bare matrix elements, and renormalized results in ratio, self-renormalization and hybrid schemes respectively in Fig.~\ref{fig:schemes_lambda_pz1}, for A-term quasi-DA of $\Lambda$ with $P^z=0.5$ GeV.




As evident from these figures, the ratio scheme results exhibit continuity and smoothness across all regions, but inevitably introduces IR effects at large distances. The self-renormalization demonstrate effective removal of UV divergences, yet exhibit singularity at short distances. The hybrid scheme, as a contrast, effectively eliminates the divergences and singularities, exhibiting smooth continuity across the regions while retaining proper normalization.

\section{Summary}\label{sec:summary}
In this report, we review the recent progress in our lattice calculations of baryon LCDAs within the LaMET framework. In particular, we successfully implement a hybrid renormalization scheme for the $\Lambda$-baryon quasi-DAs in our recent work~\cite{LatticePartonCollaborationLPC:2025vhd}. This strategy effectively removes the linear divergences inherent in lattice matrix elements, thereby enabling stable truncated Fourier transforms and well-defined perturbative matching procedures.

This work establishes an important and reliable foundation for the precise and systematically controlled determination of baryon LCDAs within the LaMET approach. As a natural extension of this study, lattice calculations of all three leading-twist LCDAs of the $\Lambda$-baryon and Proton are currently in progress and will be reported in the near future.

\end{document}